\documentclass[a4paper,11pt]{article}

\usepackage{contribution}

\newcommand{\weblink}[2][]{%
    \ifthenelse{\equal{#1}{}}%
    {\textnormal{\url{#2}}}%
    {\textnormal{\href{#2}{#1}}}%
}

\newcommand{\acknowledgements}[1]{%
  \bigskip\bigskip
  \textsf{\textbf{\Large Acknowledgements}} \\[2ex]
  {#1}
  \bigskip
}

\def\beq{\begin{equation}}
\def\eeq#1{\label{#1}\end{equation}}
\def\eeqn{\end{equation}}

\def\beqa{\begin{eqnarray}}
\def\eeqa#1{\label{#1}\end{eqnarray}}
\def\eeqan{\end{eqnarray}}

\let\bar=\overbar

\def\etal{{\it et al.}}

\def\Dslash{\not{\hbox{\kern-4pt $D$}}}
\def\dslash{\not{\hbox{\kern-2pt $\del$}}}

\def\ee{e^+e^-}

\def\msb{{\bar{\ssstyle M \kern -1pt S}}}

\newcommand{\contribution}[7][]{%
  \clearpage
  \thispagestyle{plain}
  \ifthenelse{\equal{#1}{}}
  {\hypersetup{pdftitle={#2}}}
  {\hypersetup{pdftitle={#1}}}
  \hypersetup{pdfauthor={{#3} {#4}}}
  {\centering\normalfont\LARGE\bfseries\sffamily #2 \par\nobreak}
  \lhead{}
  \chead{%
    \textit{\footnotesize XIV International Conference on Hadron Spectroscopy
      (\weblink[\textit{hadron2011}]{http://www.hadron2011.de}), 13-17 June 2011, Munich, Germany}%
  }
  \rhead{}
  \bigskip
  \begin{center}
    {#3} {#4}\ifthenelse{\equal{#6}{}}{}{\footnote{\weblink[#6]{mailto:#6}}}
    \ifthenelse{\equal{#7}{}}{}{#7} \\
    \textit{#5}
  \end{center}
  \bigskip
}

\renewcommand{\abstract}[1]{%
  \begin{center}
    \begin{minipage}{0.85\textwidth}
      \begin{footnotesize}
        #1
      \end{footnotesize}
    \end{minipage}
  \end{center}
  \bigskip
}

\begin{document}
%
%
%
%
%
{  

\newcommand{\dd}{\displaystyle}
\def\bea{\begin{eqnarray}}
\def\eea{\end{eqnarray}}
\def\be{\begin{equation}}
\def\ee{\end{equation}}
\def\nn{\nonumber}
\newcommand{\spur}[1]{\not\! #1 \,}
%

\contribution[Latest Developments in Heavy Meson Spectroscopy]
{Latest Developments in Heavy Meson Spectroscopy}  
{Fulvia}{De Fazio}  
{Istituto Nazionale di Fisica Nucleare, Sezione di Bari \\
 Via Orabona 4, I-70126 Bari, ITALY}  
{}{} 
%

\abstract{I discuss  developments in  heavy meson spectroscopy. In particular, I consider  the system of $c{\bar s}$ mesons and the puzzling state X(3872), with focus on the strategies for their  classification.}
%
\section{Introduction}
In the last decade,  many new
charm and beauty hadrons have been discovered. Some of them  fit  the quark
model scheme, others  still need to be properly classified. Here I focus on $c \bar s$  mesons and,
 to introduce the topic, I  describe the properties  of  mesons
with a single heavy quark  in the infinite
heavy quark mass limit. Then, I turn to
 the state X(3872) observed in the hidden charm spectrum.

 Before the   B-factory era, the  $c \bar s$  spectrum
consisted of
the pseudoscalar  $D_s(1968)$ and  vector $D_s^*(2112)$ mesons,
   $s$-wave states of the  quark model, and of the
axial-vector $D_{s1}(2536)$ and tensor $D_{s2}(2573)$ mesons,
$p$-wave states. In 2003,  two narrow  resonances were discovered: $D_{sJ}(2317)$ and $D_{sJ}^*(2460)$
with $J^P=0^+,\,1^+$   \cite{Aubert:2003fg,Besson:2003cp}.
 Their
identification as   $c \bar s$ states was   debated
\cite{Colangelo:2004vu};  however, they have the right quantum numbers
to complete the $p$-wave multiplet, and
their radiative decays occur accordingly, so that their
interpretation  as ordinary $c{\bar s}$ mesons seems natural and  now widely accepted \cite{Colangelo:2003vg,Colangelo:2004vu,Colangelo:2005hv}.
Afterwards,  two other $c{\bar s}$ mesons decaying to
$D K$ were observed: $D_{sJ}(2860)$  \cite{Aubert:2006mh} and $D_{sJ}(2700)$
\cite{Brodzicka:2007aa}, the latter with   $J^P=1^-$.
  Later, in \cite{Aubert:2009di} it was found that $D_{sJ}(2700)$ is  likely the first radial excitation of $D^*_s$. In \cite{Aubert:2009di} also another state was observed: $D_{sJ}(3040)$.
As discussed in Section \ref{sec4}, the predictions for the decays   of $D_{sJ}(2860)$, $D_{sJ}(2700)$ and $D_{sJ}(3040)$
 following from different
identifications can be used for the classification \cite{Colangelo:2006rq,Colangelo:2010te}.

In Section \ref{sec6}, after briefly recalling  some of the latest news in the spectroscopy of hidden charm and beauty mesons, I survey  the  properties of X(3872) and study a few radiative decay modes
which are useful to shed  light on its structure.

\section{Hadrons containing a single heavy quark $Q$}\label{sec2}

The description of mesons with a single heavy quark $Q$ is simplified in QCD in
the heavy quark  $m_Q \to \infty$ limit, when the spin $s_Q$ of the
heavy quark and  the angular momentum
 $s_\ell$ of the  light degrees of freedom: $s_\ell=s_{\bar q}+ \ell$ ($s_{\bar q}$
being the light antiquark spin and $\ell$ the orbital angular
momentum of the light degrees of freedom relative to $Q$) are
decoupled. Hence   spin-parity  $s_\ell^P$ of the light degrees of
freedom is conserved in strong interactions \cite{HQET} and mesons
can be classified as doublets of $s_\ell^P$. Two states with
$J^P=(0^-,1^-)$, denoted as $(P,P^*)$, correspond to $\ell=0$ (the {\it fundamental} doublet). The
four states corresponding to $\ell=1$ can be collected in two
doublets, $(P^*_{0},P_{1}^\prime)$ with  $ s_\ell^P={1 \over 2}^+$
and $J^P=(0^+,1^+)$, $(P_{1},P_{2})$ with $ s_\ell^P={3 \over
2}^+$ and $J^P=(1^+,2^+)$. For $\ell=2$ the doublets have
$s_\ell^P={3 \over 2}^-$, consisting of states with
$J^P=(1^-,2^-)$, or  $ s_\ell^P={5 \over 2}^-$  with
$J^P=(2^-,3^-)$ states. And so on.
For each doublet, one can consider
a tower of similar states corresponding to their radial
excitations.

One can  predict whether these states are
narrow or broad. For example, strong decays of the members of the $J^P_{s_\ell}=(1^+,2^+)_{3/2}$ doublet to the
fundamental doublet plus a light pseudoscalar
meson occur in $d$-wave. Since the rate for this process is
proportional to $|\vec p|^5$ (in general, to $|\vec p|^{2\ell+1}$,
$p$ being the light pseudoscalar momentum and $\ell$ the angular
momentum transferred in the decay), these states are expected to
be narrow. On the contrary, the members of the
$J^P_{s_\ell}=(0^+,1^+)_{1/2}$ doublet decay in $s$-wave, hence
they should be broad.

 $D_s(1968)$, $D_s^*(2112)$ belong to the lowest  $s_\ell^P={1 \over 2}^-$
doublet.
 $D_{s1}(2536)$,
$D_{s2}(2573)$ correspond to the doublet with $J^P_{s_\ell}=(1^+,2^+)_{3/2}$, $D_{sJ}(2317)$,
  $D_{sJ}^*(2460)$,  to that with $J^P_{s_\ell}=(0^+,1^+)_{1/2}$.
 Mixing between the two $1^+$ states is  allowed
 at $O(1/m_Q)$; however, for non-strange charm mesons such a mixing was
  found to be small \cite{Abe:2003zm,Colangelo:2005gb}.

In the heavy quark limit,  the various doublets are represented by
effective fields: $H_a$ for $s_\ell^P={1\over2}^-$ ($a=u,d,s$ is a
light flavour index), $S_a$ and $T_a$ for $s_\ell^P={1\over2}^+$
and $s_\ell^P={3\over2}^+$, respectively; $X_a$ and $X^\prime_a$
for
  $s_\ell^P={3\over2}^-$ and $s_\ell^P={5\over2}^-$, respectively:
 \bea
&&  \hskip 0.2cm H_a  = \frac{1+{\rlap{v}/}}{2}[P_{a\mu}^*\gamma^\mu-P_a\gamma_5]  \label{neg} \nn  \\
&&  \hskip 0.2cm S_a = \frac{1+{\rlap{v}/}}{2} \left[P_{1a}^{\prime \mu}\gamma_\mu\gamma_5-P_{0a}^*\right]   \nn \\
&&  \hskip 0.2cm T_a^\mu=\frac{1+{\rlap{v}/}}{2} \Bigg\{
P^{\mu\nu}_{2a} \gamma_\nu  -P_{1a\nu} \sqrt{3 \over 2} \gamma_5
\left[g^{\mu \nu}-{1 \over 3} \gamma^\nu (\gamma^\mu-v^\mu)
\right]
\Bigg\} \label{pos2}  \\
&&   \hskip 0.2cm X_a^\mu =\frac{1+{\rlap{v}/}}{2} \Bigg\{
P^{*\mu\nu}_{2a} \gamma_5 \gamma_\nu  -P^{*\prime}_{1a\nu} \sqrt{3 \over 2}  \left[
g^{\mu \nu}-{1 \over 3} \gamma^\nu (\gamma^\mu-v^\mu) \right]
\Bigg\}   \nn   \\
&&  \hskip 0.2cm X_a^{\prime \mu\nu} =\frac{1+{\rlap{v}/}}{2}
\Bigg\{ P^{\mu\nu\sigma}_{3a} \gamma_\sigma
 -P^{*'\alpha\beta}_{2a} \sqrt{5 \over 3}
\gamma_5 \Bigg[ g^\mu_\alpha g^\nu_\beta
-{1 \over 5} \gamma_\alpha g^\nu_\beta
(\gamma^\mu-v^\mu) -{1 \over 5} \gamma_\beta g^\mu_\alpha
(\gamma^\nu-v^\nu) \Bigg] \Bigg\} \,\,;\nn \eea the various
operators annihilate mesons of four-velocity $v$ (conserved in
strong interactions) and   contain a factor $\sqrt{m_P}$.
  At the leading order in the heavy
quark mass and light meson momentum expansion the decays  $F \to H
M$ $(F=H,S,T,X,X^\prime$ and $M$ a light pseudoscalar meson) can
be described by the Lagrangian interaction  terms (invariant under
chiral and
 heavy-quark spin-flavour transformations)
\cite{hqet_chir,Casalbuoni:1992dx}:
\bea && \hskip -0.6cm {\cal L}_H = \,  g \, Tr [{\bar H}_a H_b
\gamma_\mu \gamma_5 {\cal
A}_{ba}^\mu ] \nn \\
&& \hskip -0.6cm {\cal L}_S =\,  h \, Tr [{\bar H}_a S_b
\gamma_\mu \gamma_5 {\cal
A}_{ba}^\mu ]\, + \, h.c. \,,  \nn \\
&& \hskip -0.6cm {\cal L}_T =  {h^\prime \over
\Lambda_\chi}Tr[{\bar H}_a T^\mu_b (i D_\mu {\spur {\cal
A}}+i{\spur D} { \cal A}_\mu)_{ba} \gamma_5
] + h.c.   \label{lag-hprimo}  \\
&& \hskip -0.6cm {\cal L}_X =  {k^\prime \over
\Lambda_\chi}Tr[{\bar H}_a X^\mu_b (i D_\mu {\spur {\cal
A}}+i{\spur D} { \cal A}_\mu)_{ba} \gamma_5
] + h.c.  \nn \\
&& \hskip -0.6cm {\cal L}_{X^\prime} =  {1 \over
{\Lambda_\chi}^2}Tr[{\bar H}_a X^{\prime \mu \nu}_b [k_1 \{D_\mu,
D_\nu\} {\cal A}_\lambda   +k_2 (D_\mu D_\nu { \cal A}_\lambda + D_\nu
D_\lambda { \cal A}_\mu)]_{ba}  \gamma^\lambda \gamma_5] + h.c.
\nn \eea
where  $D_{\mu ba}
=-\delta_{ba}\partial_\mu+\frac{1}{2}\left(\xi^\dagger\partial_\mu
\xi +\xi\partial_\mu \xi^\dagger\right)_{ba}$, ${\cal A}_{\mu
ba}=\frac{i}{2}\left(\xi^\dagger\partial_\mu \xi-\xi
\partial_\mu \xi^\dagger\right)_{ba}$ and
 $\displaystyle \xi=e^{i {\cal M} \over
f_\pi}$. $\cal M$ is a matrix containing the light pseudoscalar meson fields
($f_{\pi}=132 \; $ MeV),
$\Lambda_\chi \simeq 1 \, $ GeV
 the chiral symmetry-breaking scale. ${\cal L}_S$, ${\cal L}_T$ describe
decays of positive parity heavy mesons with the emission of
light pseudoscalar mesons in $s$- and $d$- wave, respectively, $g,
h$ and $h^\prime$ representing  effective coupling constants.
 ${\cal L}_X$, ${\cal L}_{X^\prime}$  describe
the decays of  negative parity mesons with the
emission of light pseudoscalar mesons in $p$- and $f$- wave with
couplings $k^\prime$, $k_1$ and $k_2$.
The structure of the Lagrangian terms for radial excitations of
the  doublets is the same,  but  the
couplings $g, h,\dots$ have to be  substituted
by $\tilde g, \tilde h,\dots$.

\section{$c{\bar s}$ mesons: The case of $D_{sJ}(2860)$, $D_{sJ}(2700)$ and $D_{sJ}(3040)$}\label{sec4}

In 2006,  BaBar observed  a heavy $c{\bar s}$ meson,
$D_{sJ}(2860)$, decaying to $D^0 K^+$ and $D^+ K_S$, with  mass $M
= 2856.6 \pm 1.5 \pm 5.0$ MeV
and width $\Gamma = 47 \pm 7 \pm 10$ \cite{Aubert:2006mh}.
 Shortly after,
  analysing the $D^0 K^+$  invariant mass distribution in
 $B^+ \to {\bar D}^0 D^0 K^+$  Belle Collaboration
\cite{Brodzicka:2007aa}  found a $J^P=1^-$
resonance, $D_{sJ}(2710)$, with $M = 2708\pm 9
^{+11}_{-10}$  MeV and $ \Gamma= 108
\pm23 ^{+36}_{-31}$  MeV.

 In order to classify $D_{sJ}(2860)$ and $D_{sJ}(2710)$, their
strong decays were studied in \cite{Colangelo:2006rq}, comparing the predictions which follow from
different quantum number assignments.
I summarize here the main results, starting with $D_{sJ}(2860)$. A new $c \bar s$ meson decaying
to $DK$ can be either  the $J^P=1^-$ state of the  $ s_\ell^P={3
\over 2}^-$ doublet,  or the $J^P=3^-$ state of the  $ s_\ell^P={5
\over 2}^-$ one, in both cases with  lowest radial quantum number.
Otherwise $D_{sJ}(2860)$ could be a radial excitation of already
observed $c\bar s$ mesons: the first radial excitation of $D_s^*$
($J^P=1^-$   $ s_\ell^P={1 \over 2}^-$) or of $D_{sJ}(2317)$
($J^P=0^+$   $ s_\ell^P={1 \over 2}^+$) or  of $D_{s2}^*(2573)$
($J^P=2^+$ $ s_\ell^P={3 \over 2}^+$).
As for $D_{sJ}(2710)$, having $J^P=1^-$, it could be either the first radial excitation
  belonging to the $ s_\ell^P={1 \over 2}^-$
doublet  ($D_s^{* \prime}$) or
 the low lying state with $ s_\ell^P={3
\over 2}^-$ ($D_{s1}^{* }$).

For  both mesons the ratios of decay rates $R_1=
{\Gamma( D_{sJ}  \to D^*K) \over \Gamma( D_{sJ} \to DK) }$ $
R_2={\Gamma( D_{sJ} \to D_s \eta) \over \Gamma( D_{sJ} \to DK) }
$ ($D^{(*)}K=D^{(*)+} K_S +D^{(*)0} K^+$), obtained using
 eqs. (\ref{pos2}) and (\ref{lag-hprimo}),
are useful to
discriminate among the various assignments  \cite{Colangelo:2006rq}. Table
\ref{tab:ratios}  reports such ratios
in the various cases; it is interesting that
 they
 do not depend on the coupling constants,  but only on the quantum numbers.
\begin{table}[tb]
  \begin{center}
    \begin{tabular}{ l c c }
 $D_{sJ}(2860) $  &$R_1$  &   $R_2$
\\ \hline
 $s_\ell^p={1\over 2}^-$, $J^P=1^-$,  $n=2$  &$1.23$& $0.27$ \\
$s_\ell^p={1\over 2}^+$, $J^P=0^+$, $n=2$   &$0$& $0.34$ \\
$s_\ell^p={3\over 2}^+$, $J^P=2^+$, $n=2$   &$0.63$& $0.19$\\
$s_\ell^p={3\over 2}^-$, $J^P=1^-$,   $n=1$   & $0.06$& $0.23$ \\
$s_\ell^p={5\over 2}^-$, $J^P=3^-$,   $n=1$   & $0.39$& $0.13$ \\
    \hline  $D_{sJ}(2710) $  &$R_1$  &   $R_2$
\\ \hline $s_\ell^p={1\over 2}^-$, $J^P=1^-$,  $n=2$ & $0.91 $ & $0.20 $   \\
  $s_\ell^p={3\over 2}^-$, $J^P=1^-$,   $n=1$ & $0.043 $ & $0.163 $
  \\
    \end{tabular}
    \caption{Predicted ratios
    $R_1$ and $R_2$ (see text for definitions)
 for the various assignment
 of quantum numbers to  $D_{sJ}(2860) $ and $D_{sJ}(2710)$.  }
    \label{tab:ratios}
  \end{center}
\end{table}
%

I first consider
$D_{sJ}(2860)$.
The case $s_\ell^p={3\over 2}^-$, $J^P=1^-$, $n=1$ can  be
excluded since,   using $k^\prime \simeq h^\prime\simeq 0.45\pm0.05$ \cite{Colangelo:2005gb},
 would give   a width
 incompatible with the measurement.
In the  assignment  $s_\ell^p={1\over 2}^+$, $J^P=0^+$, $n=2$ the
decay to $D^*K$ is forbidden.  However, in this case $D_{sJ}(2860)$  should have  a spin partner with
 $J^P=1^+$  decaying to $D^* K$ with a small width and mass around $2860$ MeV.
To explain the absence of such a signal one should  invoke a mechanism favoring
the production of the $0^+$ $n=2$ state and inhibiting that
 of $1^+$ $n=2$ state,
 which  is difficult to imagine.

Among the remaining possibilities, the assignment
$s_\ell^p={5\over 2}^-$, $J^P=3^-$, $n=1$ seems  the most likely
one.  In this case the small $DK$ width is due to the
  kaon momentum suppression factor:
$\Gamma(D_{sJ}\to DK) \propto q_K^7$. The spin
partner, $D_{s2}^{*}$, has $s_\ell^P={5 \over 2}^-$,
$J^P=2^-$,
  decaying to $D^* K$ and not to $DK$. It
would also  be narrow in the $m_Q\to \infty$ limit,
where the transition  $D_{s2}^{*}\to D^*K$ occurs in $f$-wave. As
an effect of $1/m_Q$ corrections this  decay can occur in
$p$-wave, so that   $D_{s2}^{*}$ could be  broader; hence,  it
is not necessary to invoke a mechanism inhibiting the production
of this state with respect to $J^P=3^-$. If  $D_{sJ}(2860)$ has
$J^P=3^-$, it is not expected to  be produced
 in non leptonic $B$ decays such as
$B \to  D D_{sJ}(2860)$. Actually, in
the Dalitz plot analysis of $B^+ \to \bar D^0 D^0 K^+$ no signal of
$D_{sJ}(2860)$ was found \cite{Brodzicka:2007aa}.

In the latest  BaBar analysis  \cite{Aubert:2009di}  $D_{sJ}(2860)$ has been observed decaying to $DK$ and $D^*K$
final states, hence excluding the assignment $J^P=0^+$. However, the
measurement  \cite{Aubert:2009di} \be
{BR(D_{sJ}(2860) \to D^*K) \over BR(D_{sJ}(2860) \to DK)}=
 1.10 \pm 0.15_{stat} \pm 0.19_{syst} \nn \,\,\ee
leaves the identification of $D_{sJ}(2860)$ still an open issue.
A  confirmation  that $D_{sJ}(2860)$ is  a $J^P=3^-$
state
 could be the detection of its non-strange partner  $D_3$,   also expected to be narrow, that can  be produced in semileptonic and
in non leptonic $B$ decays
\cite{Colangelo:2000jq}.

Let us now look at
 $D_{sJ}(2710)$. As  Table \ref{tab:ratios} shows,
$R_1$ is very different if $D_{sJ}(2710)$ is $D_s^{*\prime}$ or
$D_{s1}^*$. Comparing the results in that Table with the BaBar
measurement  \cite{Aubert:2009di}: \be
{BR(D_{sJ}(2710) \to D^*K) \over BR(D_{sJ}(2710) \to
DK)}=  0.91 \pm 0.13_{stat} \pm 0.12_{syst} \nn \,\,\ee
allows to
conclude that $D_{sJ}(2710)$ is most likely $D_s^{*\prime}$, the
first radial excitation of $D_s^*(2112)$.

 From the
computed widths, assuming that  $\Gamma(D_{sJ}(2710))$ is
saturated by the considered modes and identifying $D_{sJ}(2710)$ with $D_s^{*\prime}$,  the coupling $\tilde
g$, analogous to $g$ in (\ref{lag-hprimo}) when $H$ is the doublet of the $n=2$ radial excitations, can be determined $
\tilde g= 0.26 \pm 0.05$,  a  value similar to those
obtained for analogous effective couplings  \cite{Colangelo:1995ph}.
This result for $\tilde g$  can provide information
about $D_s^\prime$, the spin partner of $D_{sJ}(2710)$ having $J^P=0^-$; it is
 the first radial excitation of $D_s$ and can  decay  to $
D^{*0} K^+$, $ D^{*+} K^0_{S(L)}$, $ D^{*}_s \eta$.
In the heavy quark limit, these partners are degenerate.  Using
the result for $\tilde g$ one predicts $
\Gamma(D_s^\prime)= (70 \pm 30)$  MeV.

Identifying $D_{sJ}(2700)$ with $D^{*\prime}_s$, its charmed  non strange partners are $D^{*\prime +}$ and $D^{*\prime 0}$,  the radial excitations of $D^{*+,0}$. Their masses can be fixed to $2600 \pm 50$ MeV assuming
that $D_{sJ}(2700)$ is heavier by an amount of the size   of the
strange quark mass.
 $D^{*\prime }$ can decay to $D^{*\prime }  \to D\pi$,
 $D_s K$, $D \eta$, $ D^{*}\pi$,  $D^{*} \eta$ so that
 the previous result for   ${\tilde g}$   gives
$\Gamma(D^{*\prime+(0)})=(128 \pm 61)$ MeV.
Noticeably,  studying $D^+ \pi^-$, $D^0 \pi^+$, $D^{*+} \pi^-$ systems,   BaBar   found four new charmed non strange mesons \cite{delAmoSanchez:2010vq} and, among these,   the state $D^*(2600)$ likely to be identified with $D^{* \prime}$ (the non strange partner of $D_{sJ}(2700)$), and the state $D(2550)^0$ likely to be the spin partner of $D^*(2600)$, corresponding  to the first radial excitation of the $D$ meson. Comparison of
 the measured widths $\Gamma(D^*(2600))=93 \pm 6 \pm 13$ MeV, $\Gamma(D(2550))=130 \pm 12 \pm 13$ MeV with
 the prediction for  $\Gamma(D^{*\prime+(0)})$ supports the proposed identification.

In  \cite{Aubert:2009di}
 another  broad structure was observed, $D_{sJ}(3040)$, with
$M= 3044 \pm 8_{stat}
(^{+30}_{-5})_{syst}$ MeV and
$\Gamma = 239 \pm 35_{stat} (^{+46}_{-42})_{syst}$
 MeV.
 $D_{sJ}(3040)$  decays to $D^*K$ and not to $DK$, hence it has unnatural parity:  $J^P=1^+, \, 2^-, \, 3^+, \cdots$.
The lightest    not yet observed states with such quantum numbers  are
the two $J^P=2^-$ states belonging to the doublets  with
$\dd s_\ell=3/2$ and  $\dd s_\ell=5/2$ denoted as $D_{s2}$ and $D_{s2}^{\prime *}$, respectively.
The identification with the radial excitations
with $n=2$,  $J^P=1^+$,  and $\dd s_\ell=1/2$ (the meson ${\tilde
D}_{s1}^\prime$) or   $\dd s_\ell=3/2$ (the meson ${\tilde D}_{s1}$)  is also
possible.
Notice that, if the identification of $D_{sJ}(2860)$ as the
$J^P_{s_\ell}=3^-_{5/2}$ meson were experimentally confirmed, this
would disfavor the assignment of $D_{sJ}(3040)$ to its spin
partner $D_{s2}^{*\prime}$ with  $J^P_{s_\ell}=2^-_{5/2}$ , since a mass inversion in a  spin doublet seems unlikely.
 For a similar reason, one would also disfavor the identification of $D_{sJ}(3040)$ with $D_{s2}$,  although in that case the two
mesons  would belong to  different doublets.
The
strong decays of $D_{sJ}(3040)$ to a charmed meson and a light
pseudoscalar one  can be evaluated using  the effective Lagrangians in Eq.(\ref{lag-hprimo}).
In particular, one can compute the ratio $R_1={\Gamma(D_{sJ}(3040) \to D_s^*
\eta) \over \Gamma(D_{sJ}(3040) \to D^* K)}$
($D^* K=D^{*0}K^+$ + $D^{*^+}K_S^0$), with results collected in Table \ref{summary} \cite{Colangelo:2010te}. The spread among them  is useful to discriminate
among the assignments, in particular between ${\tilde D}_{s1}^\prime$ and $D_{s2}^{* \prime}$.

The mass of  $D_{sJ}(3040)$ is large enough to allow decays to  $(D_0^*,D_1^\prime)K$,
$(D_1,D_2^*)K$ and $D_{s0}^*\eta$, with different
features  in the four cases.
Other   allowed modes   are   into $DK^*$ or $D_s \phi$ which can be described using an
approach  based on effective
Lagrangian terms  \cite{Casalbuoni:1992gi}.
\begin{table}[tb]
\begin{center}
\begin{tabular}{p{1.1 in} |  p{1. in} | p{1. in} |p{1. in} |p{1. in} }
   decay modes &  ${\tilde D}_{s1}^\prime$  (n=2)  & ${\tilde D}_{s1}$ (n=2)&
 $ D_{s2}$ (n=1) & $ D_{s2}^{* \prime}$ (n=1) \\
   &    $(J^P_{s_\ell}=1^+_{1/2})$  &
($J^P_{s_\ell}=1^+_{3/2})$ &
   ($J^P_{s_\ell}=2^-_{3/2})$ &  ($J^P_{s_\ell}=2^-_{5/2})$ \\
      \hline \hline
$D^* K$, $D^*_s \eta$ & $s-$ wave & $d-$ wave & $p-$ wave & $f-$
wave \\ \hline $R_1$ & 0.34 & 0.20 & 0.245 & 0.143\\ \hline \hline
$D^*_0 K$, $D^*_{s0} \eta$, $D_1^\prime K$ & $p-$ wave & $p-$ wave
& $d-$ wave & $d-$ wave \\ \hline \hline $D_1 K$ & $p-$ wave &
$p-$ wave & - & $d-$ wave \\ \hline $D_2^* K$ & $p-$ wave & $p-$
wave & $s-$ wave & $d-$ wave \\ \hline \hline $D K^*$, $D_s
\phi$ & $s-$ wave & $s-$ wave & $p-$ wave & $p-$ wave \\
\cline{2-5} &$\Gamma\simeq 140$
MeV & $\Gamma\simeq 20$ MeV  & negligible & negligible \\
   \end{tabular}
\caption{Features of the decay modes of $D_{sJ}(3040)$
 for the four proposed assignments.}\label{summary}
\end{center}
\end{table}
The results obtained  in the four possible identifications are collected in Table \ref{summary} \cite{Colangelo:2010te}, from which   some conclusions can be drawn.
The determination of the wave in which a particular decay proceeds is useful to predict
 a hierarchy among
the widths of the states in the four cases.
Consequently,   the two $J^P=1^+$ are expected to be broader than
the two $J^P=2^+$ states,   hence it is  likely that $D_{sJ}(3040)$ should
 be identified with one of such two axial-vector mesons. These can be distinguished since the widths
to the  $DK^*$ and $D_s \phi$ decay modes
 are  larger for   ${\tilde D}^\prime_{s1}$
than for ${\tilde D}_{s1}$.
 Finally, although  less probable,
the identification with  $D_{s2}$
can be discarded/confirmed studying  the $D_2^* K$ $s-$wave final
state.

\section{Heavy quarkonium and the intriguing case of X(3872)}\label{sec6}

Besides the  new charmed mesons, new heavy quarkonium or quarkonium-like states were observed. Some have been classified as standard quarkonia:
the charmonia $h_c$ \cite{Rosner:2005ry}, $\eta_c(2S)$
\cite{Choi:2002na}, $\chi_{c2}(2P)$  \cite{Uehara:2005qd}, and, in the beauty case,
the $\eta_b(1S)$ \cite{:2008vj}, $h_b(1P)$ \cite{:2011zp,Adachi:2011ji}
 and $h_b(2P)$ \cite{Adachi:2011ji}.
Others are still awaiting for the right interpretation,
since not only their quantum numbers are not  established, but
even their $Q{\bar Q}$ structure is questioned \cite{QQreview}.
Among these,  the  charged $Z(4430)^-$ state seen by Belle Collaboration in $B \to Z^- K$, decaying to $ \psi(2S) \pi^-$, $\chi_{c1}\pi^-$ \cite{:2007wga}.
The minimal quark content of this state would be $c{\bar c}u {\bar d}$, identifying it necessarily as an  exotic state.
Search for $Z^-$ was performed by BaBar, but no signal was found \cite{:2008nk}.
Later on, Belle found other charmonium-like charged  $Z$ states\cite{Mizuk:2008me}
 and, more recently, also bottomonium-like $Z_b(10610)$ and $Z_b(10650)$ states decaying to $\Upsilon(nS) \pi^\pm$ (n=1,2,3) and $h_b(mP) \pi^\pm$ (m=1,2) \cite{Collaboration:2011gja}. These states require confirmation, too.

Here I focus on  the state $X(3872)$, discovered in 2003  by Belle Collaboration  in $B^\pm \to K^\pm X \to K^\pm J/\psi \pi^+ \pi^-$ decays \cite{Choi:2003ue} and confirmed by BaBar \cite{Aubert:2004ns}, CDF \cite{Acosta:2003zx} and D0 \cite{Abazov:2004kp} Collaborations. The PDG resonance parameters  are: $M(X) = 3871.57 \pm 0.25$ MeV and $\Gamma(X) < 2.3$ MeV (90/\% C.L.) \cite{PDG}.
Looking at the $J/\psi \pi^\pm \pi^0$ channel, no charged partners
were found \cite{Aubert:2004zr}.
The mode  $X \to J/\psi \gamma$ allows to fix  charge conjugation of X to $C=+1$.
Moreover, a    $D^0 \bar D^0 \pi^0$  enhancement in
 $B \to D^0 \bar D^0 \pi^0 K$ decay
was  reported \cite{Gokhroo:2006bt} with
  $ \frac{B(X \to D^0 \bar D^0  \pi^0)}{B(X \to J/\psi \pi^+ \pi^- )}=9\pm4$,
hence $X$  mainly  decays into final states with open charm mesons.

These measurements, though not fully consistent with the
charmonium interpretation (as far as the mass of X is concerned), do not contradict it. However,
the observation of
 $X \to J/\psi \pi^+ \pi^- \pi^0$  with  the measurement $ \frac{B(X \to J/\psi \pi^+ \pi^- \pi^0)}{B(X \to J/\psi \pi^+ \pi^- )}=1.0 \pm 0.4 \pm 0.3$ \cite{Abe:2004zs}
  implies,
 considering the two modes as  induced by $\rho^0$ and $\omega$ intermediate states, isospin violation.

 The  three pion decay  is also important to fix the spin-parity of X.
While the angular analysis  in $X \to J/\psi \pi^+ \pi^-$  favours  $J^P=1^+$,    studies of the three pion distribution in  $X \to J/\psi \omega \to J/\psi \pi \pi \pi$ are more favourable to  $J^P=2^-$ \cite{delAmoSanchez:2010jr}.
Hence, if X is a $c{\bar c}$ state it can be  either  the first radial excitation of $\chi_{c1}$, $\chi_{c1}^\prime$,  or  the state $\eta_{c2}$ having $J^{PC}=2^{-+}$.

On the other hand, the peculiar features of X  suggested the conjecture
 that  it is not a charmonium state. In particular,
the   coincidence between its mass   and the
 $D^{*0} \overline D^0$ mass:
 $M(D^{*0} \overline D^0)=3871.2\pm 1.0$ MeV,   inspired  the proposal that
$X(3872)$  could be  a molecule \cite{okun}, a bound state of  $D^{*0}$ and $\overline D^0$
with small binding energy \cite{Voloshin:2007dx},  an interpretation that would  account for a few
properties of $X(3872)$. For example,
 if
 the wave function of $X(3872)$  has various hadronic components \cite{voloshin1}
one could explain why this state seems not to have  definite isospin. However,   the molecular
binding mechanism still needs to be clearly identified, while
 the description of X(3872) as a charmonium state presents alternative arguments to the
molecular description \cite{Barnes:2003vb,charmonium}.  Concerning
the isospin  violation, to correctly
interpret the large  ratio $ \frac{B(X \to J/\psi \pi^+ \pi^- \pi^0)}{B(X \to J/\psi \pi^+ \pi^- )}$ one
has to consider that phase space effects in two and three pion
modes are very different and it turns out that the isospin violating amplitude is
$20\%$ of the isospin conserving one \cite{suzuki}:
$ \frac{B(X \to J/\psi \rho^0)}{B(X \to J/\psi \omega )}\simeq 0.2$.

 I focus on two studies of X decays.
The first one \cite{Colangelo:2007ph} compares the charmonium versus the molecular interpretation, discussing the argument
that, if X(3872) is a $DD^*$ molecule the  decay  $X \to D^0 {\bar D}^0 \gamma$ should be dominant with respect to
$X \to D^+ D^- \gamma$,  such decays
being mainly due to the decays of its meson components \cite{voloshin1}.
In order to discuss whether this is true, in \cite{Colangelo:2007ph} the ratio $R={\Gamma(X \to D^+ D^- \gamma)\over \Gamma(X \to D^0 {\bar D}^0 \gamma)}$ has been computed assuming that X(3872) is an ordinary
$J^{PC} = 1^{++}$ charmonium state.

\begin{figure}[htb]
  \begin{center}
   \includegraphics[width=0.32\textwidth] {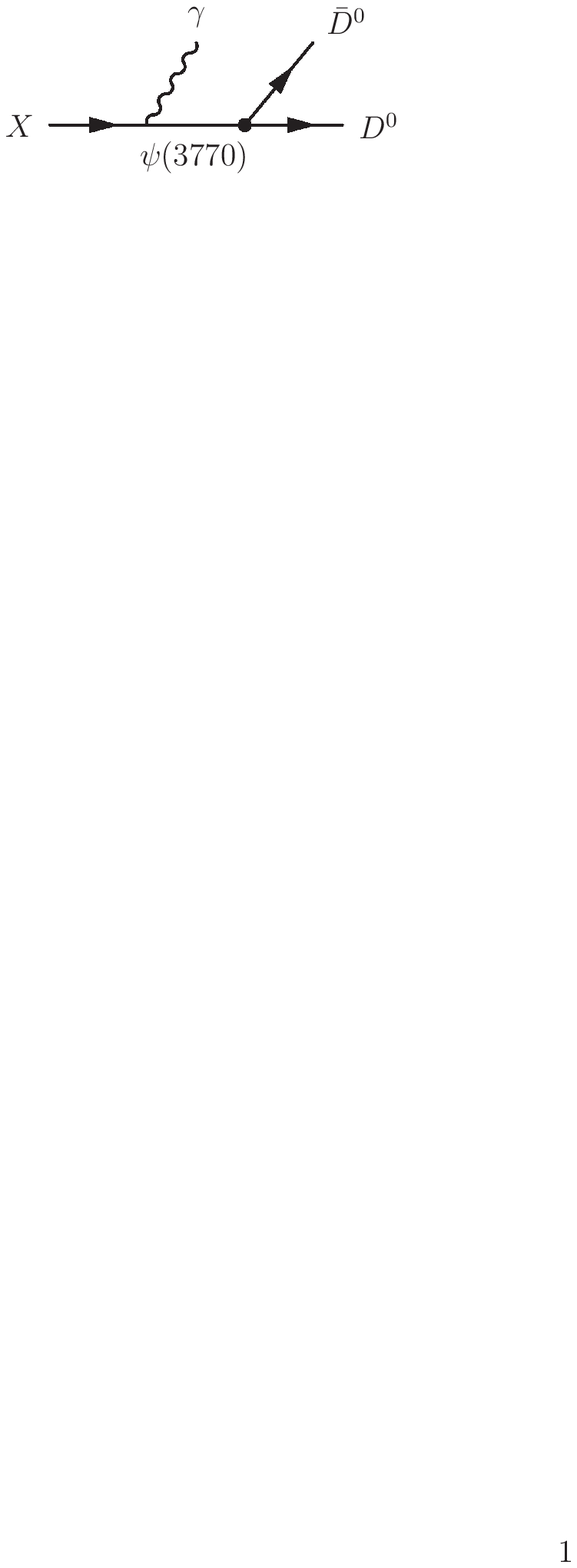}
\includegraphics[width=0.32\textwidth] {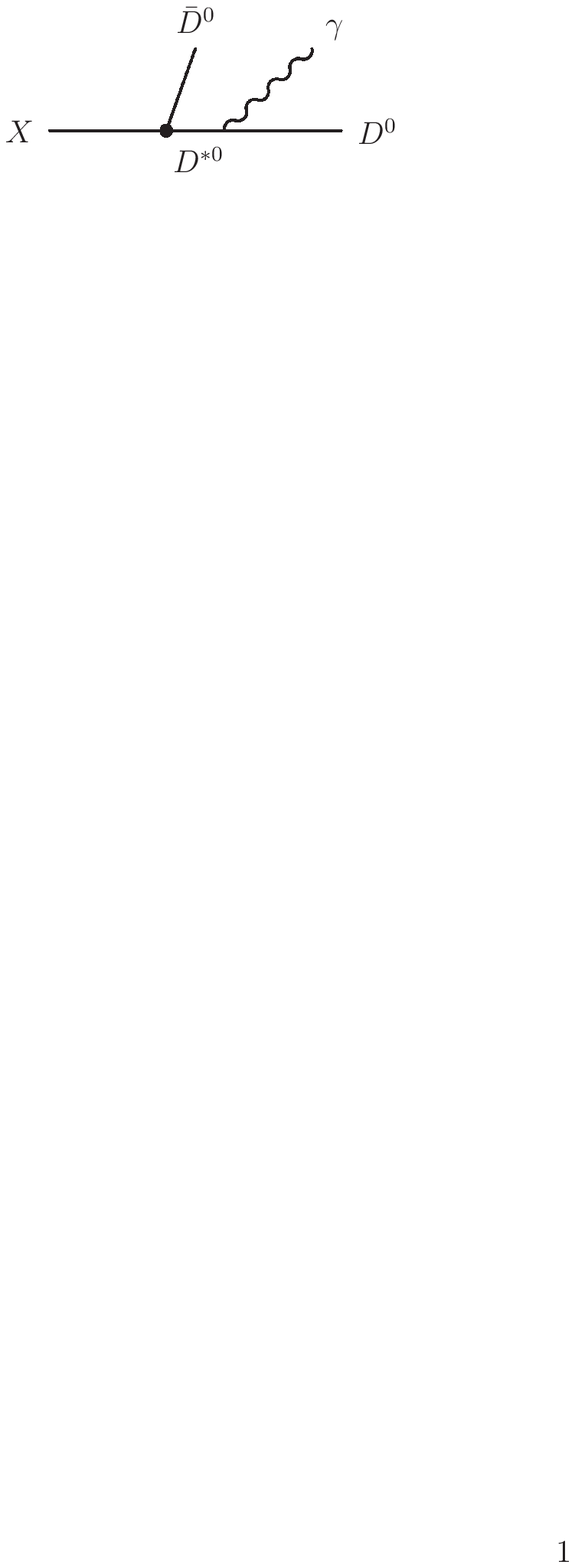}
\includegraphics[width=0.32\textwidth] {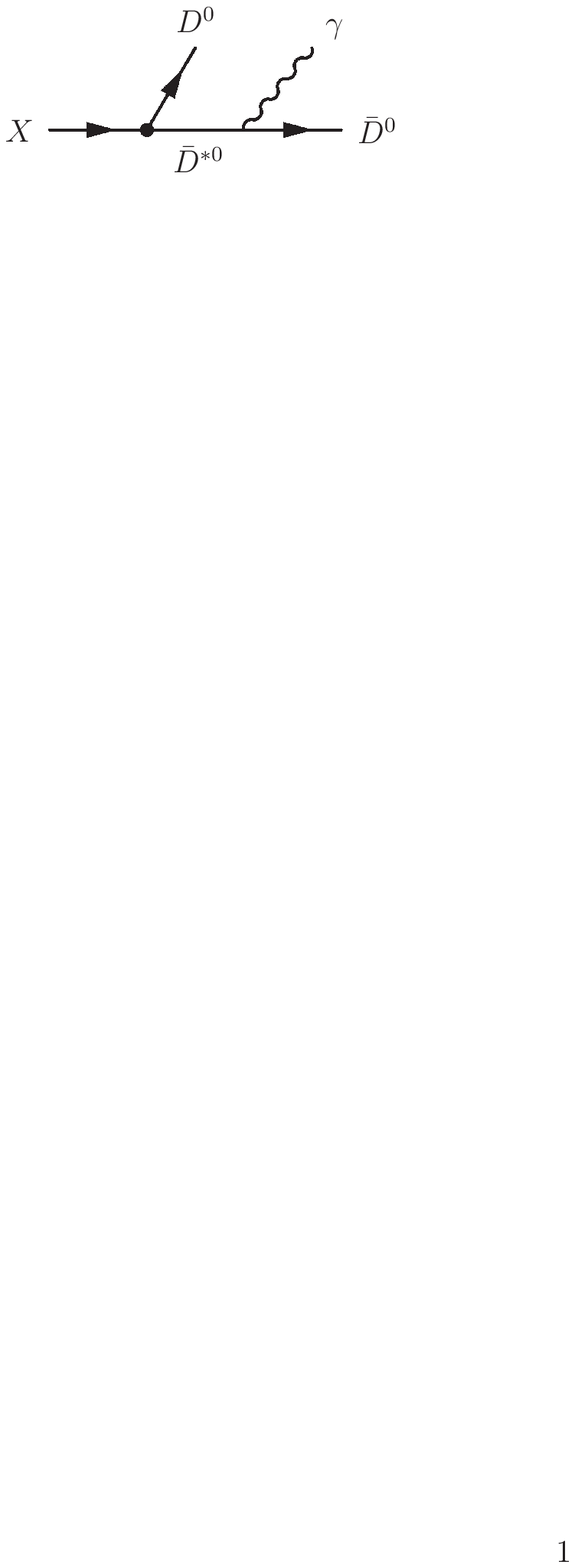}\vspace*{-11.5cm}
   \caption{Diagrams describing the radiative  modes $X \to D \bar D \gamma$.}
  \label{fig:gamma}
  \end{center}
\end{figure}
%
The transition $X(3872)\to D
\bar D \gamma$
can be studied  assuming that the
radiative decay amplitude is dominated by polar diagrams with $D^*$  and the
$\psi(3770)$  mesons as
intermediate states nearest to their mass shell (fig.\ref{fig:gamma}). These
amplitudes can be expressed in terms of two  unknown  quantities:
 the coupling constant ${\hat g}_1$ governing the $X \bar D D^*(D \bar D^*)$ matrix elements,  and the
 one appearing in the $X \psi(3770) \gamma$ matrix element.
 For  the matrix element  $X \bar D D^*(D \bar D^*)$  one can  use  a
formalism suitable to describe the interaction of the heavy
charmonium with  the doublet $H$ in (\ref{pos2}) \cite{Casalbuoni:1996pg}.
In  the
multiplet: \be P^{(Q \bar Q)\mu}=\left( {1 + \spur{v} \over 2}
\right) \left( \chi_2^{\mu \alpha}\gamma_\alpha +{1 \over
\sqrt{2}}\epsilon^{\mu \alpha \beta \gamma} v_\alpha \gamma_\beta
\chi_{1 \gamma}+ {1 \over \sqrt{3}}(\gamma^\mu-v^\mu) \chi_0
+h_1^\mu \gamma_5 \right)\left( {1 - \spur{v} \over 2} \right) \hspace{-0.2cm}
\label{pwave} \ee
the fields $\chi_2$, $\chi_1$, $\chi_0$ correspond
to the spin triplet with $J^{PC}=2^{++}, 1^{++}$, $0^{++}$, respectively,
while the spin singlet $h_1$ has $J^{PC}=1^{+-}$.  If
$X(3872)=\chi_{c1}^\prime$, it is described by $\chi_1$.
The strong interaction with the $D$ and $D^*$ mesons can be
described by the effective Lagrangian \cite{pham}
\be {\cal L}_1= i {g_1} Tr
\left[P^{(Q \bar Q)\mu} {\bar H}_{1a} \gamma_\mu {\bar H}_{2a}
\right] + h.c. \,\,.\label{lagr1new} \ee
Using  (\ref{lagr1new})
the couplings $XDD^*$
 which enter in the  second and the
third diagrams in fig.\ref{fig:gamma}, can be expressed in terms of
the   dimensionless coupling constant $\hat g_1=g_1 \sqrt{m_D}$.
Notice that,
due to  isospin symmetry, the couplings of the meson
$X$ to charged and neutral $D$ are equal, at odds with the molecular description where
$X$ mainly  couples to neutral $D$.

The
matrix element
$<D(k_1)  \gamma(k,{\tilde \epsilon})|D^{*}(p_1,\xi)> =i \, e \, c^\prime \,
\epsilon^{\alpha \beta \tau \theta} \, {\tilde \epsilon}^*_\alpha
\, \xi_\beta \, p_{1 \tau} \, k_{\theta}$ is also required.
The parameter $c^\prime$ accounts for the coupling
of  the photon  to  both the charm and the light quark  and can be fixed from data on radiative $D^{*+}$ decays
\cite{PDG}.

To compute the
first diagram in fig.\ref{fig:gamma}  the matrix element
$<\psi_{(3770)}(q , \eta) \gamma(k, \tilde \epsilon)|X(p,
\epsilon)>=i \, e \,c \, \epsilon^{\alpha \beta \mu \nu} \, {\tilde
\epsilon}^*_\alpha \, \epsilon_\beta \, \eta^*_\mu \, k_\nu $ is needed;
$c$   is an unknown parameter.
On the other hand,   the coupling
$\psi(3770) D {\bar D}$  can be fixed  from  experiment to
$g_{\psi D \bar D}=25.7 \pm 1.5$.

Putting all the ingredients together one obtains  the
ratio $ R={\Gamma(X \to D^+ D^- \gamma) \over
\Gamma(X \to D^0 \overline D^0 \gamma)}$, plotted in fig.\ref{fig:ratio} \cite{Colangelo:2007ph} versus   ${c \over {\hat
g}_1}$ ,
showing that  the radiative $X$ decay into charged
$D$ mesons  is always suppressed with respect to the mode with
neutral $D$ and in any case
$R < 0.7$.  Moreover, for small values of $ {c \over \hat g_1}$  the ratio $R$ is tiny, so that
this is not peculiar of a molecular structure of $X(3872)$.
\begin{figure}[htb]
  \begin{center}
    \includegraphics[width=0.4\textwidth]{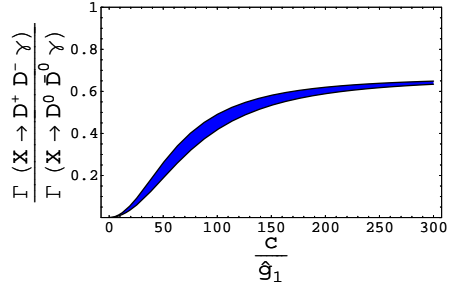}
    \caption{\footnotesize{Ratio of  $X \to D^+ D^- \gamma$ to
 $X \to D^0 \bar D^0 \gamma$ decay  widths  versus the ratio
of  parameters $c/\hat g_1$.}}\label{fig:ratio}
  \end{center}
\end{figure}
%

${\hat g}_1$ enters also in  the mode $X(3872)\to
D^0\bar D^0 \pi^0$   that can be considered  as induced by intermediate $D^*$ states. The amplitude depends on
the
coupling constant $D^*D\pi$, proportional to the constant  $g$
 in eq. (\ref{lag-hprimo}).
  Using data on $D^{*+}$ decays to $D\pi$ \cite{PDG}, one can derive
$g=0.64 \pm 0.07$.
This   allows to constrain ${\hat g}_1<4.5$ from the upper bound  $\Gamma(X\to D^0 {\bar D}^0 \pi^0) < \Gamma(X(3872))<2.3 \,$ MeV.
Hence, a value of ${\hat g}_1$
of the typical size of the hadronic couplings can reproduce the small width of $X(3872)$.

The second analysis that I discuss also aims at shedding light on the structure of X(3872) through the calculation of its radiative decay rates to $J/\psi \gamma$ and $\psi(2S) \gamma$ assuming that it is the state $\chi^\prime_{c1}$ \cite{DeFazio:2008xq} and using an effective Lagrangian
approach which exploits spin symmetry for heavy $Q{\bar Q}$ states
\cite{Casalbuoni:1992yd}.
Unlike the heavy-light $Q{\bar q}$ mesons, in heavy quarkonia
there is no heavy flavour symmetry \cite{Thacker:1990bm}, hence it would not be  possible to exploit data on
charmonium to obtain quantitative information on bottomonium or
viceversa. However,  at a qualitative level,
bottomonium system can help in understanding charmonium.

A heavy $Q{\bar Q}$ state ($Q=c,\,b$) can be identified by $n^{2s+1}L_J$  as a meson  with parity $P=(-1)^{L+1}$
and charge-conjugation $C=(-1)^{L+s}$:
$n$ is  the radial
quantum number, $L$ the orbital angular momentum, $s$ the spin
and  $J$ the total angular momentum.
Radiative transitions
between states belonging to the same $nL$ multiplet to states
belonging to another $n^\prime L^\prime$ one are described in terms of a single
coupling constant $\delta^{nLn^\prime L^\prime}$.

I introduce the effective fields for the states involved in the  decays  $X \to J/\psi \gamma$ and $X \to \psi(2S) \gamma$. Identifying X with the state $\chi^\prime_{c1}$, it belongs to the multiplet with $L=1$   introduced in (\ref{pwave}). $J/\psi$ and $\psi(2S)$ are described by the $J^P=1^-$ $H_1$ component of the  doublet:
\be
J={ 1+ \spur{v} \over 2} \left[H_1^\mu \gamma_\mu -H_0 \gamma_5
\right]{ 1- \spur{v} \over 2} \,\,.\label{Swave} \ee

 The effective Lagrangian describing
radiative transitions among members of the $P$ wave  and of the
$S$ wave multiplets has been derived in \cite{Casalbuoni:1992yd}:
\be {\cal L}_{nP \leftrightarrow mS}=\delta^{nPmS}_Q Tr
\left[{\bar J}(mS) J_\mu(nP) \right] v_\nu F^{\mu \nu} + \rm{h.c.}
\,.\label{lagPS} \ee  $F^{\mu \nu}$ the electromagnetic field
strength tensor.
Hence,  a single constant $\delta^{nPmS}_Q$
describes all the transitions among the members of the $nP$
multiplet and those of the $mS$ one.

I consider  the ratios $ R_J^{(b)}={\Gamma(\chi_{bJ}(2P) \to \Upsilon(2S) \, \gamma )
\over \Gamma(\chi_{bJ}(2P) \to \Upsilon(1S) \, \gamma )}$,  proportional to $R_\delta^{(b)}={\delta_b^{2P1S} \over
\delta_b^{2P2S}}$ $(J=0,1,2)$. From the measured
branching ratios of  $\chi_{bJ}(2P)\to \Upsilon(1S) \, \gamma \, ,\Upsilon(2S) \, \gamma$  \cite{PDG}, the
average value  can be obtained: $ R_\delta^{(b)}=8.8 \pm 0.7 $. It is
reasonable that, even though the couplings might be different
in the beauty and the charm cases, their ratios  stay stable. Therefore, using the result for $ R_\delta^{(b)}$ in the case of $\chi_{c1}^\prime$ decays, I get:
\be R_1^{(c)}={\Gamma(\chi_{c1}(2P) \to \psi(2S) \,
\gamma ) \over \Gamma(\chi_{c1}(2P) \to \psi(1S) \, \gamma )}=1.64
\pm 0.25 \label{ratioXth} \,.\ee

In
\cite{:2008rn} the following ratio has been measured \footnote{Belle Collaboration has recently provided an upper limit for the Ratio $R_X<2.1$ (at 90\% C.L.) \cite{Bhardwaj:2011dj}.}:
 \be R_X={\Gamma(X(3872) \to \psi(2S) \, \gamma ) \over
\Gamma(X(3872) \to \psi(1S) \, \gamma )}=3.5 \pm 1.4
\label{ratioX} \,.\ee
 In view of the underlying
approximation, one can conclude that the experimental
value in (\ref{ratioX}) and the theoretical prediction (\ref{ratioXth}) are close enough to
consider plausible
the identification $X(3872)=\chi_{c1}(2P)$, in contrast  to
the composite scenarios, in which the mode  $X(3872) \to \psi(2S)
\, \gamma$ is suppressed compared to $X(3872) \to
\psi(1S) \, \gamma$ \cite{Barnes:2003vb,Swanson:2006st}.

\section{Conclusions}\label{concl}

In the last decade,   many predicted charm and beauty mesons have been discovered, along with many unexpected ones.
In the case of $D_{sJ}$ mesons, the   analysis of their decay modes  allows to classify them as ordinary $c{\bar s}$ states, although the identification of $D_{sJ}(2860)$ is still  under scrutiny.

The case of hidden charm and beauty mesons is more complicated. As for $X(3872)$,  two analyses of the radiative decays of X show that the charmonium interpretation seems to be a likely one, although experimentally it is still unclear whether its spin-parity is $J^P=1^+$ or $J^P=2^-$.

\acknowledgements{I thank P. Colangelo, R. Ferrandes, S. Nicotri, A. Ozpineci and M. Rizzi for collaboration.}


%

}  


\end{document}